\documentclass[aps,prb,floatfix,twocolumn,showpacs,superscriptaddress]{revtex4}%\documentclass[aps,prl,twocolumn,showpacs,preprintnumbers,amsmath,amssymb,superscriptaddress]{revtex4}%\usepackage{here}\usepackage{graphicx}\usepackage[ngerman, english]{babel}%\newcommand{\figref}[1]{Fig.\,\protect\ref{#1}}\newcommand{\missing}[1]{\textcolor{red}{\textbf{Missing: #1}}} \makeatletter\def\ScaleIfNeeded{%\ifdim\Gin@nat@width>\linewidth \linewidth \else \Gin@nat@width \fi
\makeatother

\makeatletter
\def\ScaleIfNeeded{%
\ifdim\Gin@nat@width>\linewidth \linewidth \else \Gin@nat@width \fi
} \makeatother
\usepackage{amsmath}
\usepackage{amssymb}
\usepackage{graphicx}% Include figure files
\usepackage{longtable}%
\usepackage{dcolumn}% Align table columns on decimal point
\usepackage{bm}% bold math
\usepackage[usenames]{color}%font colors
\usepackage{ulem}
\usepackage{subfigure}
\usepackage{color}
\newcommand{\red}[1]{\textcolor{black}{#1}} %question

\begin{document}
%\preprint{}
\title{Solitonic spin-liquid state due to the violation of the Lifshitz condition in Fe$_{1+y}$Te}
\author{Ph. Materne}
\affiliation{Institut f\"ur Festk\"orperphysik, Technische
Universit\"at Dresden, 01062 Dresden, Germany}%%
\author{C. Koz}
\affiliation{Max Planck Institute for Chemical Physics of Solids,
N\"othnitzer Stra\ss e 40, 01187 Dresden, Germany}%
\author{U. K. R\"o{\ss}ler}
\affiliation{IFW Dresden, Postfach 270016, 01171 Dresden, Germany}
\author{M. Doerr}
\affiliation{Institut f\"ur Festk\"orperphysik, Technische Universit\"at Dresden,
01062 Dresden, Germany}
\author{T. Goltz}
\affiliation{Institut f\"ur Festk\"orperphysik, Technische Universit\"at Dresden,
01062 Dresden, Germany}%
\author{H. H. Klauss}
\affiliation{Institut f\"ur Festk\"orperphysik, Technische Universit\"at Dresden,
01062 Dresden, Germany}%
\author{U. Schwarz}
\affiliation{Max Planck Institute for Chemical Physics of Solids,
N\"othnitzer Stra\ss e 40, 01187 Dresden, Germany}
\author{S. Wirth}\affiliation{Max Planck Institute for Chemical Physics of Solids,
N\"othnitzer Stra\ss e 40, 01187 Dresden, Germany}%
\author{S. R\"o{\ss}ler}
\email{roessler@cpfs.mpg.de}\affiliation{Max Planck Institute for Chemical Physics of Solids,
N\"othnitzer Stra\ss e 40, 01187 Dresden, Germany}%%\date{\today}
\begin{abstract}
A combination of phenomenological analysis and M\"ossbauer spectroscopy experiments
on the tetragonal Fe$_{1+y}$Te system
% , parent to the `11-type' Fe-based superconductors,
indicates that the magnetic ordering transition
in compounds with higher Fe-excess, $y\ge$~0.11,
is unconventional.
Experimentally, a liquid-like magnetic precursor with quasi-static
spin-order is found from significantly broadened M\"ossbauer spectra
at temperatures  above the antiferromagnetic transition.
The incommensurate spin-density wave (SDW) order in Fe$_{1+y}$Te is described by a magnetic free energy
that violates the weak Lifshitz condition in the Landau theory of second-order transitions.
The presence of multiple Lifshitz invariants provides the mechanism
to create multidimensional, twisted, and modulated solitonic phases.
\end{abstract}%

\maketitle%\textbf{insert introduction}

Complex ordering in solids often is associated with precursor phenomena. In the non-ordered state,  some features of the true long-range order (LRO) are already present locally or temporarily. The onset of these precursors above the transition usually is not distinguished by a clear thermodynamic anomaly. Some examples are the normal state (pseudogap) in doped cuprates \cite{Lee} and heavy fermion superconductors,\cite{PhysRevLett.100.137003} magnetic polarons in the colossal magnetoresistive manganites \cite{salamon} and EuB$_{6},$~\cite{zhang} blue phases in chiral nematic liquid crystals, \cite{Wright89} and precursor phenomena in non-centrosymmetric helimagnets\cite{Pappas2009, Wilhelm2011,Barla2015}.

Here, we argue that similar precursor phenomena as in
the non-centrosymmetric liquid-crystals and magnets
exist in a great number of other condensed matter systems.
In particular, magnetic ordering into modulated states
\red{provides} a rich variety of systems where such effects
should be observable \cite{Dzyaloshinskii64,ToledanoToledano1987,Bogdanov1989}.
Phenomenological Landau theory is a secure guide
to identify candidate systems for such effects.
If the \red{order parameter} can be rotated by the presence of several twisting terms, known
as Lifshitz invariants, in the phenomenological free-(Landau-Ginzburg)-energy,
a continuous phase transition into a homogeneously ordered phase is impossible \cite{Michelson1978}.
Ordering under the influence of such twisting
terms is described by \textit{Dzyaloshinskii models} \cite{Dzyaloshinskii64}.
Corresponding systems can display liquid-like mesophases as precursors, which are composed
of {\textit{multidimensional}}, localized (solitonic) states\cite{Bogdanov1995} that can be fluctuatingly nucleated
before LRO is established \cite{Pappas2009, Wilhelm2011}.

In this study, we identify one interesting example for such a \red{magnetically} modulated precursor.
The tetragonal Fe$_{1+y}$Te ( y = 0.13 and 0.15) is a parent
to one class of Fe-based superconductors which undergoes a magnetic ordering
transition at low temperatures into a modulated antiferromagnetic state \cite{rodriguez,roessler2011,Zaliznyak2012,koz2013,Rodriguez2013} which has been described as continuous. However, it does not obey the standard Landau-theory.
Using symmetry arguments, we have constructed the corresponding Dzyaloshinskii
theory and show that it violates the weak Lifshitz criterion\cite{Michelson1978}.
This motivates the search for a magnetic precursor state similar to those
found in chiral helimagnets \cite{Pappas2009,Wilhelm2011,Barla2015}.
The Fe-based material is ideally suited for identifying such quasi-static magnetic states
above the long-range ordering temperature, as the hyperfine-field distribution observable in the M\"ossbauer spectroscopy reflects the inhomogeneous spin-state
established in these liquid-like spin-structures. This precursor state signals
an exotic magnetic ordering in Fe-chalcogenides \red{within} an intermediate temperature range.

The  non-superconducting parent compounds of  Fe-based superconductors are interesting
systems on their own, as they show a complex interplay of magnetic and structural phase
transitions, which indicate a competition of several ordering phenomena. Among the different families of Fe-based superconductors, the chalcogenides with '11'-structure can be considered as reference systems owing to their representative and simple crystal structure.
%Although the crystal structure of Fe$_{1+y}$Te closely resembles other Fe-based
%superconductors, neutron scattering studies \cite{Bao, Li, Igor, rodriguez} on Fe$_{1+y}$Te
%identified some peculiar properties. The observed ordered moment is significantly large
%with an unconventional temperature enhanced magnetism\cite{Igor}.
%
Fe$_{1+y}$Te has a particularly rich phase-diagram \cite{roessler2011,koz2013,Rodriguez2013,Koz2012}
\red{and} neutron scattering studies \cite{Bao, Li, Igor, rodriguez, Fobes14, Stock2014} on these materials have
identified some peculiar properties.
%The magnetic ordering transition of Fe$_{1+y}$Te
%has been thoroughly investigated in experiments \cite{Bao, Li, Igor, rodriguez}.
The propagation vector along ($0,~\pi$) at low Fe-excess suggests that nesting does
not play a crucial role (see, e.g. Ref.~\onlinecite{Ding13}), but demonstrates the
existence of a microscopic mechanism for the formation of an antiferromagnetic type of order
similar to a \red{spin-density wave} (SDW). The observed ordered moment in Fe$_{1+y}$Te is significantly large in comparison to \red{the} parent pnictides.
 Fe$_{1+y}$Te also displays \red{an unconventional, temperature-enhanced} magnetism\cite{Igor}. The low-temperature crystal and magnetic structures,
as well as the commensurability of the propagation vector in Fe$_{1+y}$Te depend on the occupancy of the excess Fe ($y$)
in the crystallographic 2$c$-site within the $P$4$/$nmm space group.  Compounds of higher doping $y\ge 0.11$ undergo
a continuous transition \cite{roessler2011, Zaliznyak2012} into an incommensurate, and possibly helical, magnetic state \cite{Ducatman14}.
The onset \red{temperature} of this LRO increases with Fe content: $T_N(y=0.11)=57$~K, and $T_N(y=0.15)=63$~K. For $0.11 < y \le 0.13$,
this is followed by a transition into a collinear SDW phase by a first-order magneto-structural phase transition,
i.e., a lock-in transition takes place at \red{a} temperature $T_l < T_N$ \cite{roessler2011}.

The experimentally observed continuous magnetic ordering into an incommensurate modulated state in Fe$_{1+y}$Te for $y\ge 0.11$
must obey a generalized (non-local) Landau theory \cite{ToledanoToledano1987}.
Considering the propagation vector $\mathbf{q} \sim (0,\Delta,1/2)$ with $\Delta\simeq$ 0.38 to 0.45 reported for  Fe$_{1+y}$Te with  $y \ge 0.11$\cite{Bao, rodriguez}, the ordering modes are SDWs with sets of propagation vectors that have the same symmetry as $\mathbf{q}$, around a base mode $\Phi$\red{,} with propagation vector $\mathbf{q}_0$ \cite{Dzyaloshinskii64}.
The free energy density is expanded in these $\mathbf{q}$-depending $\Phi$ and its gradients.
The leading square term in order parameter (OP) components is
\begin{equation}
\label{w_L}%
w_L=\Phi^{*}\,[\alpha(T-T_0) + \delta (\red{\mathbf{k}-\mathbf{q}_0}) +\omega(\red{\mathbf{k}-\mathbf{q}_0})^2 + \dots ]\,\Phi,
\end{equation}
where the function $\Phi$ transforms according to an irreducible (co)representation induced by the little group of the bare propagation vector $\mathbf{q}_0$ and a summation over propagation vectors \red{$\mathbf{k}-\mathbf{q}_0$} is implied.
The experimentally observed $\mathbf{q}$ lies on a line of symmetry\cite{supplement}.
Thus, the irreducible representations (irreps) of the little group belonging to $\mathbf{q}$ \red{and labeled} U$_i$, $i=1,\dots,4$, have one degree of freedom, $\nu=1$.
Eventually, a commensurate order can be reached by a lock-in transition with propagation vector $\mathbf{q}_1 = (0, 1/2, 1/2)$. In this state, the order is described by irreps R$_j$, $j=1,2$, which are the degenerated \red{versions} of the 4 modes  U$_i$ owing to higher symmetry.
The observation of propagation vector $\mathbf{q}$ implies that the magnetically ordered state should be described by (\ref{w_L}) with a small $\delta$.
The symmetry analysis for this incommensurate magnetic order belonging to irreps U$_i$ (and also for irreps R$_j$) in Fe$_{1+y}$Te shows that there are $\mu > 1$ Lifshitz-type invariants, $i.e.$, antisymmetric terms \(\Phi_{i}\bar{\partial}_{\xi}\Phi_{j} \equiv \Phi_{i}{\partial}_{\xi}\Phi_{j} - \Phi_{j} {\partial}_{\xi} \Phi_{i}\) that couple different OP components\cite{ISOTROPY}.
%The free energy density contains a number  $\mu> 1$ of different Lifshitz invariants, i.e., antisymmetric linear gradient terms, \(\Phi_{i}\bar{\partial}_{\xi}\Phi_{j} \equiv \Phi_{i}{\partial}_{\xi}\Phi_{j} - \Phi_{j} {\partial}_{\xi} \Phi_{i}\) that couple components of the OP in different spatial directions $\xi$.
% Because $\mu > \nu$ the `weak Lifshitz condition' for continuous transitions into modulated states is violated \cite{Michelson1978}.
%
%The symmetry analysis for this  incommensurate magnetic order belonging to irreps U$_i$ (and also for irreps R$_j$) in Fe$_{1+y}$Te shows that there are $\mu>1$ Lifshitz-type invariants that couple different OP components \cite{ISOTROPY}.

The free energy expansion (\ref{w_L}) can be extended to a phenomenological Dzyaloshinskii model by collecting the low-order invariant terms in OP components.  Any magnetic ordering based on these modes must be unconventional also for the case of a mixed order that involves several propagation vectors as proposed, e.g., in terms of a plaquette ordering pattern \cite{Ducatman2012,Ena2014}.  However, the exposition here will be given by considering only one arm with a $\mathbf{q}=(0,q,1/2)$ r.\ l.\ u.\  from the star of four propagation vectors belonging to modes with label U$_i$. This restriction is a reasonable simplification because the incommensurable magnetic order in Fe$_{1+y}$Te for $y \geq 0.11$ is linked to an orthorhombic distortion owing to strong spin-lattice couplings \cite{Bao}. Thus, the modulation direction is determined by a strong anisotropy, and magnetic states as superposition of states with several propagation vector directions appear unlikely to occur.
The \red{spin density} is described by \(\mathbf{S}^{(\pm)} (\mathbf{r})=(\mathbf{a}^{(+)} \,\cos[\mathbf{q}_0\cdot\mathbf{r}],\mathbf{a}^{(-)}\,\sin[\mathbf{q}_0\cdot\mathbf{r}])\), with the local polarization given by three-component vectors $\mathbf{a}^{(\pm)}$ slowly varying on the scale of the lattice parameter.  The free energy density for a modulated state of this type (in Cartesian coordinate system with directions $a, b, c$ parallel to the lattice vectors of the tetragonal crystal, and the propagation vector along $b$)
\begin{eqnarray}
 \label{f}
 f  & =&  A [(\nabla \mathbf{a}^{(+)})^2 +   (\nabla \mathbf{a}^{(-)})^2 ] \\
  &\ & + \sum_{i=a,b,c}  B_i [( {\partial}_i {a}^{(+)}_i)^2+
                         ( {\partial}_i {a}^{(-)}_i)^2 ]
+ G \,\mathbf{a}^{(+)} \cdot \bar{\partial}_b \mathbf{a}^{(-)}  \, \nonumber \\
&\ & + \sum_{k=a,c} \left\{   D_k  [a^{(+)}_k \bar{\partial}_k a^{(+)}_b \right. + a^{(-)}_k \bar{\partial}_k a^{(-)}_b]  \nonumber \\
&\ &    \hspace{36pt}  +  E_k  [a^{(+)}_k \bar{\partial}_k a^{(-)}_b - \left. a^{(-)}_k \bar{\partial}_k a^{(+)}_b] \right\} \nonumber \\
&\ & + \alpha (T-T_0)\,  [  (\mathbf{a}^{(+)})^2 + (\mathbf{a}^{(-)})^2 ]
  + \beta [(\mathbf{a}^{(+)})^2 + \mathbf{a}^{(-)})^2]^2   \nonumber \\
&\ &  +\sum_{i=a,b,c}  \left\{  \kappa_i  [a^{(+)}_i+a^{(-)}_c]^2  \right. +\left. \gamma_i [a^{(+)}_i-a^{(-)}_c]^2 \right\} \, \nonumber ,
\end{eqnarray}
%
%\begin{eqnarray}
% \label{f}
%
% f   =  A [(\nabla \mathbf{a}^{(+)})^2 \!\!\!&+&\!\!\!   (\nabla \mathbf{a}^{(-)})^2 ] \\
 % + \sum_{i=a,b,c}  B_i [( {\partial}_i {a}^{(+)}_i)^2+
%                         ( {\partial}_i {a}^{(-)}_i)^2 ]
%\!\!\!&+&\!\!\!
 %   G \,\mathbf{a}^{(+)} \cdot \bar{\partial}_b \mathbf{a}^{(-)}  \, \nonumber \\
%+ \sum_{k=a,c} \left\{   D_k  [s^{(+)}_k \bar{\partial}_k s^{(+)}_b \right. \!\!\!&+&\!\!\! s^{(-)}_k \bar{\partial}_k s^{(-)}_b]  \nonumber \\
 %    +  E_k  [s^{(+)}_k \bar{\partial}_k s^{(-)}_b \!\!\!&-&\!\!\! \left. s^{(-)}_k \bar{\partial}_k s^{(+)}_b] \right\} \nonumber \\
 %+ \alpha (T-T_0)\,  [  (\mathbf{a}^{(+)})^2 + (\mathbf{a}^{(-)})^2 ]
  %\!\!\!&+&\!\!\! \beta [(\mathbf{a}^{(+)})^2 + \mathbf{a}^{(-)})^2]^2   \nonumber \\
 %+\sum_{i=a,b,c}  \left\{  \kappa_i  [s^{(+)}_i+s^{(-)}_c]^2  \right. \!\!\!&+&\!\!\!\left. \gamma_i [s^{(+)}_i-s^{(-)}_c]^2 \right\} \, \nonumber ,
%\end{eqnarray}
%
describes isotropic and anisotropic exchange by coefficients $A, B_i$.
The anisotropic terms are due to relativistic spin-orbit effects and should be weak.
The next terms are $\mu=5$ Lifshitz-type invariants.
Importantly, the term with coefficient $G$ is related to the presence of the incommensurate base magnetic mode.
Invariants of this type originate from spin-exchange \cite{Dzyaloshinskii64, ToledanoToledano1987}.
They are not necessarily weak compared to the exchange term $A$.
However, these terms cause a shift of the \red{wave vector} from $\mathbf{q_0}$ to $\mathbf{q}$.
Therefore, in Fe$_{1+y}$Te it must be weak enough so that a modulated state with propagation vector $\mathbf{q}$ can be realized and the expansion according to (\ref{w_L}) remains valid.
The presence of such a term is corroborated by a strong temperature dependence of the magnitude \red{$|\mathbf{q}|$} of the effective propagation vector, in agreement with the experimental observations on Fe$_{1+y}$Te \cite{Bao, rodriguez, stock2011, Par2012}.
The next terms with coefficients $D_k, E_k$ originate from relativistic spin-orbit effects
and are weak. These Dzyaloshinskii-Moriya (DM) couplings become operative because the incommensurate modulation breaks inversion symmetry and lowers the point group-symmetry of the crystal from 4$/$mmm (D$_{4h}$) to  2mm (C$_{2v}$).
The remaining homogeneous invariants correspond to the Landau expansion with bare transition temperature $T_0$, and \red{the} mode coupling term $\beta$.
\red{Additional} leading anisotropic terms with coefficients $\kappa_i, \gamma_i$ are included to signal
their crucial role for the thermodynamic stability of \red{the} modulated states and the lock-in transition.
The explicit Landau-Ginzburg free energy demonstrates that a continuous phase transition according
to Landau theory cannot take place in Fe$_{1+y}$Te for higher Fe contents where experiments observed
a continuous transition into an incommensurate helix-like antiferromagnetic order\cite{rodriguez,roessler2011,Zaliznyak2012,koz2013,Rodriguez2013,stock2011,Par2012}.
The conundrum must be resolved by a closer experimental search for a precursor state above $T_{N}$.
%XXXXXXXXX
%
\begin{figure}[t!]
	\centering
		\includegraphics[width=9 cm,clip]{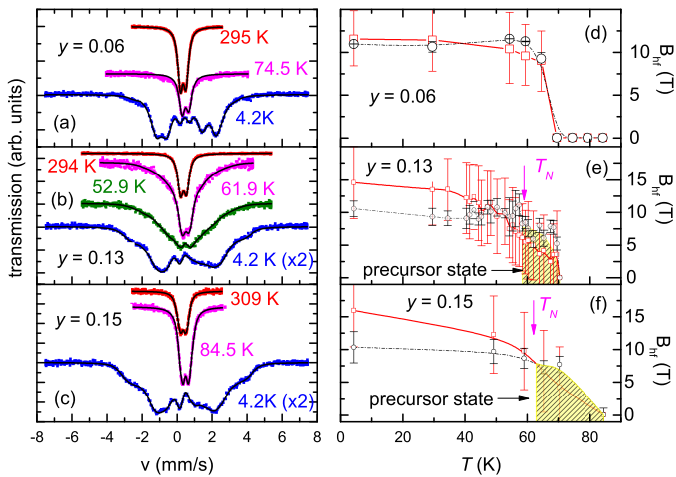}
	\caption{M\"ossbauer spectra for  Fe$_{1+y}$Te and fits (black curves) for $y=$~0.06 (a), 0.13 (b), and 0.15 (c). Corresponding $T$-dependence of the magnetic hyperfine field $B_{\rm hf}$ for $y=$~0.06 (d), 0.13 (e), and 0.15 (f). Black circles and dashed lines correspond to the values obtained for pattern 2 ($cf.$ text) while red lines and squares correspond to the average value of the magnetic hyperfine field distribution, which were obtained by \red{the} maximum entropy method (the error bars show standard \red{deviations}). Lines are \red{guides to the eye}.}
	\label{fig:MB_Bhf}
\end{figure}

For this purpose, we employ $^{57}$Fe M\"ossbauer spectroscopy on polycrystalline powder samples of Fe$_{1.06}$Te, Fe$_{1.13}$Te, and  Fe$_{1.15}$Te. These compositions were chosen because they fall in three different regimes of the phase diagram\cite{roessler2011, rodriguez}.
%For the experiments, polycrystalline samples were synthesized by solid state reaction of Fe (Alfa Aesar, 99.995\%) and Te pieces (Chempur, 99.9999\%) in vacuum sealed quartz  tubes. The details of the sample preparation can be found elsewhere.\cite{Koz2012} The samples were characterized by powder x-ray diffraction, wave length dispersive x-ray (WDX) analysis, magnetization, resistivity, and specific heat measurements. These studies ensured homogeneity of all the studied samples within the experimentally detectable limit. About 50 mg of powdered sample was taken for the M\"ossbauer measurements.
The $^{57}$Fe M\"ossbauer spectra were recorded in the temperature range \red{$T =$ 4.2 -- 300~K} using a standard M\"ossbauer setup. $^{57}$Co in a Rh matrix was used as a room temperature $\gamma$-ray source.
The M\"ossbauer spectra were analyzed by diagonalizing the hyperfine Hamiltonian including electric quadrupole and magnetic hyperfine interactions combined with the maximum entropy method (MEM) in thin absorber approximation{\cite{supplement, MSlineWidth}}.
The experimentally observed M\"ossbauer spectra for selected temperatures are given in \red{Figs.} 1~(a)-(c).
The spectra in the paramagnetic temperature region were analyzed using two patterns (subscripts 1 and 2).
Each pattern contains a set of hyperfine parameters, namely the \textit{z}-component of the electric field gradient (EFG) $V_{zz}$, the isomer shift $\delta$ (with respect to $\alpha$-Fe) and \red{the} intensity \textit{I}.
In the magnetic temperature region, pattern 1 was described \red{by} a magnetic field distribution, which was obtained by \red{MEM,} while pattern 2 was described by a Lorentzian \red{sextet} pattern with a Gaussian-distributed magnetic field.
The obtained relative \red{intensities} of pattern 1 (2) \red{are} 0.91 (0.09) for $y =$~0.06, 0.85 (0.15) for $y =$~0.13, and 0.77 (0.23) for $y =$~0.15.
Therefore, it appears reasonable to assign pattern 1 to iron on the tetragonal sites and pattern 2 to the excess iron (also see Ref.~\onlinecite{supplement}).
The relative intensities were fixed over the hole temperature range.
It turns \red{out} that both isomer shifts $\delta_{1,2}$ are of \red{approximately} equal value 0.472(5)~mm/s, 0.482(5)~mm/s and 0.479(5)~mm/s within error bars for $y =$~0.06, 0.13, and 0.15, respectively, at room temperature.
With decreasing temperature, the isomer shifts increase to values of $\delta_{1,2}^{y=0.06} \approx~0.63(1)$~mm/s, while for $y =$~0.13 and 0.15 a splitting is visible, as $\delta_1^{y=0.13}~=0.69(3)$~mm/s and $\delta_1^{y=0.15}~=0.67(3)$~mm/s are smaller than $\delta_2^{y=0.13}~=0.79(2)$~mm/s and $\delta_2^{y=0.15}~=0.74(3)$~mm/s at 4.2~K, respectively.
At room temperature, the obtained values of $V_{zz,1}$ are 17.2(6)~V/\r{A}$^2$, 18.4(1)~V/\r{A}$^2$, and 17.2(6)~V/\r{A}$^2$, while the second pattern has values of $V_{zz,2}$~=~32(1)~V/\r{A}$^2$, 38(1)~V/\r{A}$^2$, and 36(2)~V/\r{A}$^2$ for $y =$~0.06, 0.13, and 0.15, respectively.
With decreasing temperature, $V_{zz,1}$ increases slightly to values of 19.6(3)~V/\r{A}$^2$, 21.0(3)~V/\r{A}$^2$, and 20.5(5)~V/\r{A}$^2$, while $V_{zz,2}$ increases to 40(1)~V/\r{A}$^2$, 39.3(9)~V/\r{A}$^2$, and 40(1)~V/\r{A}$^2$ for $y =$~0.06, 0.13, and 0.15 at $T~\approx~80$~K, respectively.
The \textit{z}-component of the EFG was fixed for temperatures below 80~K.
For Fe$_{1.06}$Te, the onset of the magnetic ordering is indicated by the appearance of a broadened sextet structure below 70~K.
An abrupt onset of static magnetism can be seen from the temperature dependence of magnetic hyperfine field \red{Fig.} \ref{fig:MB_Bhf}(d).
This behavior confirms the first-order magneto-structural transition in Fe$_{1.06}$Te.
The magnetic field distribution of pattern 1 shows a single peak\cite{supplement}, which is consistent with a commensurate \red{SDW} and in good agreement \red{with} neutron diffraction measurements\cite{rodriguez}.
For \red{samples} Fe$_{1.13}$Te and Fe$_{1.15}$Te,
the temperature dependence of both the average value of this hyperfine field distribution as well as that obtained from pattern 2 are shown in \red{Fig.} \ref{fig:MB_Bhf}(e) and (f).
The corresponding field distributions\cite{supplement} show a rectangular shape consistent with Ref. \onlinecite{Blachowski2012}.
For the Fe$_{1.13}$Te sample, a broadening of the spectral line-width was already observed at 70.4~K indicating a quasi-static magnetic signal above $T_N~=~$57~K.
For Fe$_{1.15}$Te, a similar behavior is observed. $i.e.$, a quasi-static magnetic signal is found starting from \textit{T}$\approx 80$~K, which is clearly above $T_N~=~$62~K.
Upon increasing the amount of excess iron, the low-temperature magnitude of the magnetic hyperfine field remains constant for the excess iron while the average value of the hyperfine field distribution increases.
%
%XXXXXXXXX
\renewcommand{\floatpagefraction}{1.0}
\renewcommand{\textfraction}{0.0}
\renewcommand{\topfraction}{1.0}
\renewcommand{\bottomfraction}{1.0}
\begin{figure}[ht]
\includegraphics[width=7.2 cm,clip]{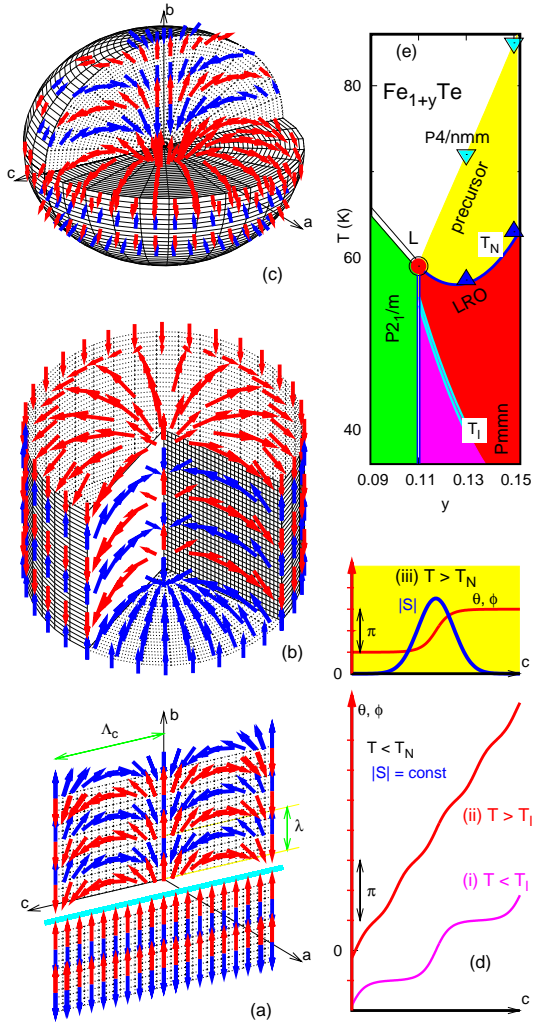}
\caption{(a)--(c) Magnetic states described by the phenomenological Dzyaloshinskii model, Eq.~(\ref{f}). (The staggered atomic spin-structure in $c$-direction is represented by one sublattice only for clarity.) (a) A one-dimensional cycloidal spiral transverse (shown arbitrarily in $c$
direction) to the propagating incommensurable spin-density wave along $b$ (upper half) represents the LRO state. Lower half shows the commensurate collinear SDW-state. (b) Two-dimensional skyrmionic texture with a double-twisted core in the $ac$-plane. (c) Three-dimensional solitonic state.
(d) Profiles of one-dimensional states of  modulated, $s_b \propto
|S|\,\exp(i\phi)$, or twisted form, $(s_b,s_c) \propto
|S|\,(\cos(\theta),\sin(\theta))$: (i)  Commensurate states with $|S|$ solitons
below the lock-in transition.
(ii) Incommensurate LRO. (iii) Isolated amplitude-modulated solitons that arise as fluctuations in the precursor state. (e) Schematic phase diagram for Fe$_{1+y}$Te. A spin-liquid state of solitons like those shown in (b) and (c) with amplitude-modulation (d(iii)) precede the incommensurate SDW-like state. Triangles  mark experimental transition temperatures for the onset of the precursor  and of $T_N$. At $T_l$ a lock-in \red{into} a commensurate state occurs. Point 'L'
is a \red{tricritical} Lifshitz point of the spin system associated with  structural phase transitions.
}
%\caption{(i) -- (iii) Magnetic states described by the phenomenological Dzyaloshinskii model, Eq.~(\ref{f}). (The staggered atomic spin-structure in $c$-direction is represented by one sublattice only for clarity.) (i) A one-dimensional cycloidal spiral transverse (shown arbitrarily in $c$ direction) to the propagating incommensurable spin-density wave along $b$ (upper half) represents the LRO state. Lower half shows the commensurate collinear SDW-state. (ii) Two-dimensional skyrmionic texture with a double-twisted core in the $ac$-plane. (iii) Three-dimensional solitonic state.
%(iv) Profiles of one-dimensional states of  modulated, $s_b \propto |S|\,\exp(i\phi)$, or twisted form, $(s_b,s_c) \propto |S|\,(\cos(\theta),\sin(\theta))$: (a)  Commensurate states with solitons below the lock-in transition.
%(b) Incommensurate LRO. (c) Isolated amplitude-modulated state. (v) Schematic phase diagram for Fe$_{1+y}$Te. A spin-liquid state of solitons like (ii) and (iii) precedes the incommensurate SDW-like state. Triangles mark experimental transition temperatures for the onset of the precursor and of $T_N$. At $T_l$ a lock-in to a commensurate state occurs. Point 'L' is a tricricital Lifshitz point of the spin-system associated with structural phase transitions.
%}
\label{States}
\end{figure}
%\clearpage

The salient point of the M\"ossbauer experiments is the observation of
a quasi-static state at temperatures above the onset of LRO for the samples
with $y\ge 0.11$ in Fe$_{1+y}$Te. The exact nature of this precursor state
would require a quantitative treatment of the various parameters
entering the Dzyaloshinskii model, Eq.~(\ref{f}). The required detailed
assessment of spin-relativistic couplings and anisotropy effects
is not available for Fe$_{1+y}$Te at present, and only qualitative
statements can be made \red{based upon} the Dzyaloshinskii model.
This model describes equilibrium magnetic states as minima of the magnetic free energy (\ref{f}),
which contain various types of modulated and localized magnetic states \cite{Dzyaloshinskii64, Bogdanov1989, Bogdanov1995, Roessler2006, Leonov201012}.
Depending on the character and strengths of the magnetic anisotropies, $\kappa_i$, $\gamma_i$ and $B_i$, and the DM-couplings,
the basic magnetic modulation with propagation in $b$-direction is twisted into one-dimensional long-period spirals propagating within the $ac$-plane.  The pitch of these spiral modulations is given by $\Lambda_k  \sim ( A + B_k)/(2\,D_k)$ for propagation in $k=a,c$ directions. The spin-structure in this state is modulated in two spatial directions with a short period, $\lambda = 2\pi/q$, of the SDW in $b$-direction, while the polarization direction slowly twists over lengths $\Lambda \gg \lambda $ in a transverse direction, Fig.~\ref{States}(a).
This magnetic state can undergo a lock-in transition into a collinear magnetic structure, if the anisotropies become strong enough, see profiles in Fig.~\ref{States}(d)(i)-(ii). The twisting of the basic collinear SDW transverse to the propagation vector $\mathbf{q}$ yields an interpretation for \red{those} experimental results \red{which} describe
helical or three-dimensional spin-structures in Fe-chalcogenides as inferred from neutron diffraction\cite{Bao,rodriguez}.
Simultaneous twisting in $a$ and $c$ direction driven by the terms $D_a$ and $D_c$ describes the \textit{\red{double-twisted}} cores of a chiral skyrmionic texture, Fig.~\ref{States}(b), as earlier calculated for systems with C$_{nv}$ symmetry \cite{Bogdanov1989, Bogdanov1994, Bogdanov2002}. However, the double-twisted configuration of $\mathbf{a}^{\pm}$ is not topologically stable. The skyrmionic strings display the  ``Alice behavior'' \cite{Sch1982,Leo2000}, as they can be annihilated by a 180-degree phase  shift of the primary modulation.
A numerical analysis of two-dimensional textures, as described by (\ref{f}) in the $ac$-plane, revealed that the exchange anisotropy terms can stabilize a staggered assembly of half-skyrmions, representing defective solitonic textures at temperatures above the onset of the one-dimensional spiral configuration \cite{Roessler2006,Leonov201012}. Such textures are inhomogeneous, the amplitude $|\mathbf{S}|$ of the local OP varies strongly and passes through zero \cite{Janson2014,Yamashita87,Mukamel1985,Roessler2011}. The temperature interval for these precursors is set by the magnitude of the Lifshitz invariants and must be sizable for Fe$_{1+y}$Te ($y\ge 0.11$) because of the exchange term $G$.

The most intriguing states described by (\ref{f}) are 3-dimensional solitons, Fig.~\ref{States}(c).
The cores of the  multidimensional states, Fig.~\ref{States}(b) and (c), can
be considered as local superpositions of the one-dimensional profiles shown in Fig.~\ref{States}(d)(iii):
A modulation in propagation direction $b$, driven by the term with coefficient $G$,
is combined with a double-twisting of the OP in transverse directions. These states
are localized. Close to and above $T_N$,
their amplitude may decay towards zero in the three spatial directions, Fig.~\ref{States}(c).

Thermal and quenched disorder impede the formation of long-range ordered assemblies of
such 2-dimensional or 3-dimensional solitons. Hence, liquid-like precursor states are formed.
Multidimensional solitonic liquid-like states can \red{more easily} adjust to random collective pinning and are favored compared to long-period spirals
for larger Fe-excess. The phase diagram of Fe$_{1+y}$Te in Fig.~\ref{States}(e) locates the liquid-like precursor state, as found from the experiment. The diagram displays the expected qualitative features, in particular an enlarged temperature range for the precursor state with increasing $y$
and a tricritical Lifshitz point (L).  \textit{Dynamic} precursor fluctuations may explain observations of unconventional
magnetic fluctuations in  Fe$_{1+y}$Te with $y<$~0.11 by neutron scattering \cite{Igor,stock2011,Par2012}.

In conclusion, the incommensurate SDW order in  Fe$_{1+y}$Te embodies elementary mechanisms to generate long-period, twisted, and multidimensionally modulated solitonic condensates. The experimental observation of a quasi-static magnetic order above $T_N$ by M{\"o}ssbauer spectroscopy confirms the existence of a liquid-like precursor state.  The stabilizing mechanism of these textures is expected to be realized in various other systems with incommensurate basic ordering modes, if they support multiple Lifshitz-type invariants. As the polarization of magnetic ordering vectors usually is rotatable and may belong to modes of different symmetry, spin-density wave systems are natural candidates for such behavior. Similar exotic behavior is also expected in magneto-electric or magneto-elastic systems allowing for a coupling of an incommensurate mode with other secondary \red{order parameters}, because Lifshitz-type invariants do exist for almost all propagation vectors in the Brillouin zone with exception of special symmetry points\cite{ToledanoToledano1987,Stokes1984}.

The authors thank A.~N.~Bogdanov, Yu.~Grin, O.~Stockert, and L.~H.~Tjeng
for stimulating discussions.
Financial support from Deutsche Forschungsgemeinschaft (within the programs SPP1458 and GRK1621) is gratefully acknowledged.
%
%TTTTTTTTTTTTTTTTTTTTTTTTTTTTTTTTTTTTTTTTTTTTTTTTTTTTTTTTT
%


\begin{thebibliography}{999}
\bibitem{MSlineWidth}The natural line width of the 57-Fe transition is 0.097 mm/s. A new radioactive 57-Co source usually has a line width of 0.11-0.12 mm/s which increases slowly to higher values of  about 0.15 mm/s after some time. As the source was not new, the line width of 0.13 mm/s is considered to be related to the source and confirms a homogeneous behavior of our samples.
\end{thebibliography}
\end{document}